\documentclass[10pt]{iopart} 

\usepackage{graphicx}
\usepackage{dcolumn}
\usepackage{bm}
\usepackage{color}
\newcommand{\add}[1]{\textcolor{black}{#1}}
\newcommand{\Tczero}{$T_{\mathrm{c}}^{\mathrm{zero}}$}
\newcommand{\Tconset}{$T_{\mathrm{c}}^{\mathrm{onset}}$ }
\newcommand{\Tc}{$T_{\mathrm{c}}$ }

\newcommand{\C}{$^\circ \mathrm{C}$}
\begin{document}

\title[]{Interface superconductivity in FeSe thin films on SrTiO$\mathrm{_3}$ grown by the PLD technique}

\author{T Kobayashi, H Ogawa, F Nabeshima, and A Maeda}

\address{Dept. of Basic Science, University of Tokyo, Meguro, 153-8902, Tokyo, Japan}
\ead{kobayashi-tomoki375@g.ecc.u-tokyo.ac.jp}

\begin{abstract}
In this study, we fabricate 5--30-nm-thick films of FeSe/STO using pulsed laser deposition (PLD).
The grown films exhibit superconductivity with an onset \Tc that is much higher than that of bulk FeSe under ambient pressure.
The observed \Tc values are exceptionally high in terms of the
strain vs. \Tc relationship of the same material established so far.
Furthermore, \Tc increases as the film thickness decreases, except for films thinner than 10 nm.
This thickness dependence of \Tc is in good agreement with the results reported for films grown by molecular beam epitaxy (MBE) that exhibit interface superconductivity.
These results indicate the realization of interface superconductivity in PLD-grown FeSe/STO.
Furthermore, our PLD technique requires no post-annealing to realize interface superconductivity, which is different from MBE techniques.
Because the PLD technique has the advantage that various interfaces can be fabricated easily by simply altering the target materials,
our results open novel routes to study interface superconductivity toward higher \Tc by systematic control of the interface.
\end{abstract}

%
\vspace{2pc}

%
\maketitle
%
\ioptwocol

\section{Introduction}
The iron chalcogenide superconductor FeSe has garnered significant attention for its high tunability of the superconducting transition temperature ($T_\mathrm{c}$).
Although the onset \Tc ($T_\mathrm{c}^\mathrm{onset}$) of bulk FeSe is 9 K at ambient pressure \cite{Hsu_2008}, it is highly enhanced by various methods.
For instance, \Tconset increases to 38 K by applying hydrostatic pressure \cite{Mizuguchi2008,Sun2016}.
The doping of electrons by intercalation \cite{Guo_2010,Shi_2018} or the electric field effect \cite{Lei_2016,Shiogai_2016,Wang_2016,Hanzawa_2016,Kouno_2018, Shikama_2020,Shikama_2021} increases \Tconset to 50 K.
In particular, zero resistivity below 46 K is realized in  \cite{Shikama_2020,Shikama_2021}.
Furthermore, angle-resolved photoemission spectroscopy (ARPES) measurements reported that a monolayer FeSe film on $\mathrm{SrTiO_{3}}$ (STO) \cite{Wang_2012,He_2013} exhibited gap opening at the Fermi level below 65 K, which has been interpreted as a manifestation of superconductivity.
The \Tc enhancement in monolayer FeSe on STO (FeSe/STO) is attributed to interfacial effects, such as electron doping from the STO substrate\cite{Liu_2012} and electron-phonon interaction with phonons of STO \cite{Lee_2014}.
Thus, the superconductivity in FeSe/STO is expected to be the first interface-enhanced superconductivity, where the existence of the interface plays a crucial role \cite{Ginzburg_1964,Allender_1973}.

In contrast to the ARPES results, resistivity measurements reported a \Tconset of 40--45 K \cite{Wang_2012, Pedersen_2020, Faeth_2021} and a much lower zero-resistance temperature (\Tczero$\!$), and these are lower than those of electron-doped FeSe systems \cite{Shikama_2020,Shikama_2021}.
This discrepancy in the reported values of \Tc observed by different measurement techniques suggests that the exact feature of interface superconductivity in FeSe/STO is still far from being fully understood.
Therefore, a more detailed investigation of the interfacial effects is essential.
In this context, the artificial control of the interface is a promising approach.
For feasible control of the interface, pulsed laser deposition (PLD)  has advantages in fabricating various heterostructures at low cost because different materials can be deposited by simply switching the target materials.
However, all studies on the interface superconductivity of FeSe published so far were performed using the molecular beam epitaxy (MBE) technique.
This is because it is difficult to grow high-quality thin films of compounds that contain highly volatile Se using other growth techniques.
Indeed, none of the previous studies by PLD succeeded in fabricating FeSe films that exhibited the interface superconductivity. 

In this paper, we report the successful growth of  FeSe/STO which exhibits interface superconductivity, using the PLD technique for the first time.
We fabricate 5--30-nm-thick FeSe films on atomically flat STO substrates.
The grown films exhibit a \Tconset higher than 19 K, which is much higher than that of bulk FeSe \cite{Hsu_2008}.
The enhanced \Tc cannot be explained in terms of the in-plane strain effect on the bulk superconductivity of FeSe \cite{Feng_2018, Nabeshima_2018,Ghini_2021,Nakajima_2021}.
Except for films thinner than 10 nm, \Tc increases with decreasing film thickness. This thickness dependence of \Tc is in good agreement with those reported for MBE-grown FeSe/STO \cite{Wang_2015, Zhao_2018}.
These results suggest that we realized interface superconductivity in FeSe/STO, grown using the PLD technique for the first time.

\section{Experimental methods}
\begin{figure}[htbp]
\includegraphics[width=\linewidth]{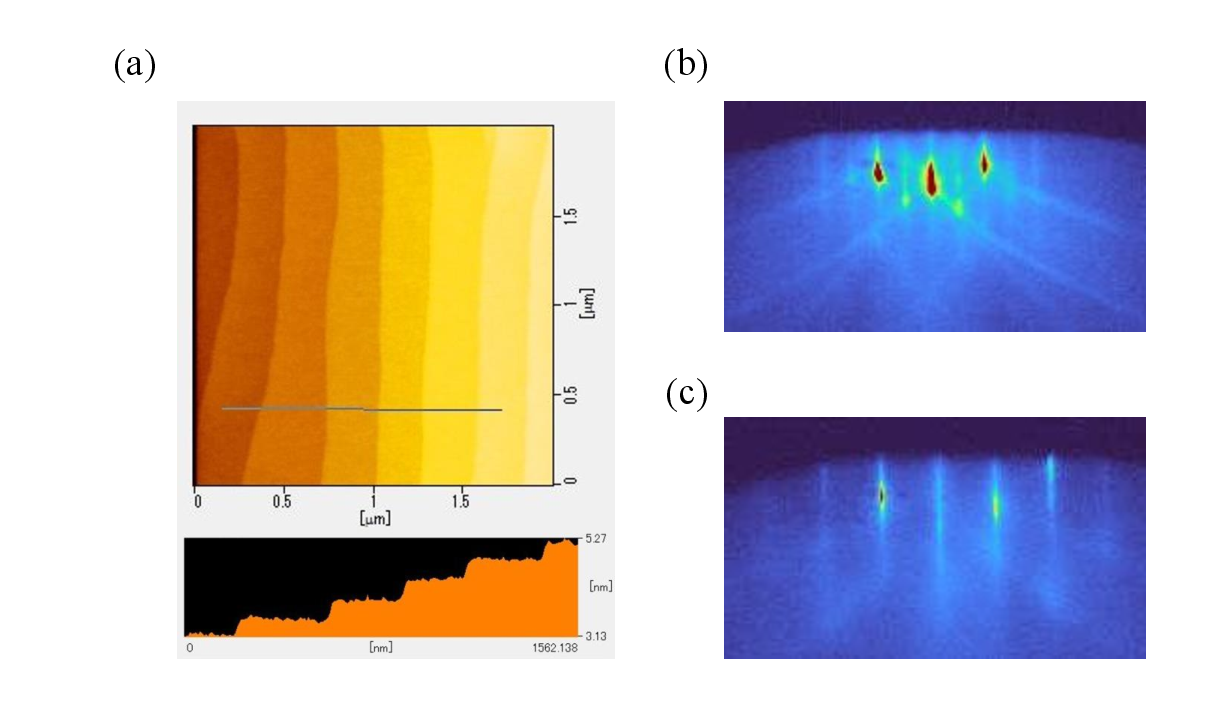}
\caption{(a) Atomic force microscope topographic image of an STO substrate after the surface treatment.  Clear step terrace structures are observed. (b) RHEED image of an STO substrate just before the deposition. (c) RHEED image of a grown FeSe film.}
\label{fig:STOsub} 
\end{figure}

All the FeSe thin films in this study were grown on insulating STO (001) substrates using PLD with a KrF laser.
The STO substrate was annealed at 1000\(^\circ\)C in air and rinsed with water to obtain a $\mathrm{TiO_2}$-terminated surface with a step-terrace structure \cite{Connell_2012} (figure \ref{fig:STOsub}(a)), which is essential to realize interface superconductivity in FeSe/STO using PLD.
Prior to deposition, the substrate was then kept for 2 hours at a growth temperature of 500\C.
Reflection high-energy electron diffraction (RHEED) images of the substrates exhibited the ($2 \times 1$) reconstructed patterns \cite{Wang_2012} under vaccum at the growth temperature.
Films were deposited by ablating the $\mathrm{Fe_{1.1}Se}$ target \cite{Feng_2018} at a growth rate of 0.2--0.5 nm/min. The laser repetition rate was 1--5Hz and the target-to-substrate distance was 50 mm.
The RHEED patterns of the grown films displayed streaks, indicating a flat surface of the films (figure \ref{fig:STOsub}(c)).
Because RHEED oscillation in FeSe/STO was sometimes not observed in this study, the RHEED intensity of the films on LaAlO$_{3}$, which were grown simultaneously as the films on STO, was observed during the deposition to control the film thickness.

The orientation and crystal structure of the grown films were characterized using X-ray diffraction (XRD) measurements with Cu K$\alpha$ radiation at room temperature.
The \textit{c}--axis and \textit{a}--axis lengths of the films were estimated from (00\textit{l}) (\textit{l} = 1, 2, 3, and 4), (\add{$\pm$}204)\add{, and (0$\pm$24)} reflections \add{using 2$\theta$-$\omega$ scans}.
The thicknesses of the films were estimated from the Laue fringe around the (001) reflection peak or using a Dektak 6M stylus profiler.
Resistivity was measured using a physical property measurement system (PPMS), from 2 to 200 K.
\section{Results and discussion}
\begin{figure}[htb]

\includegraphics[width=\linewidth]{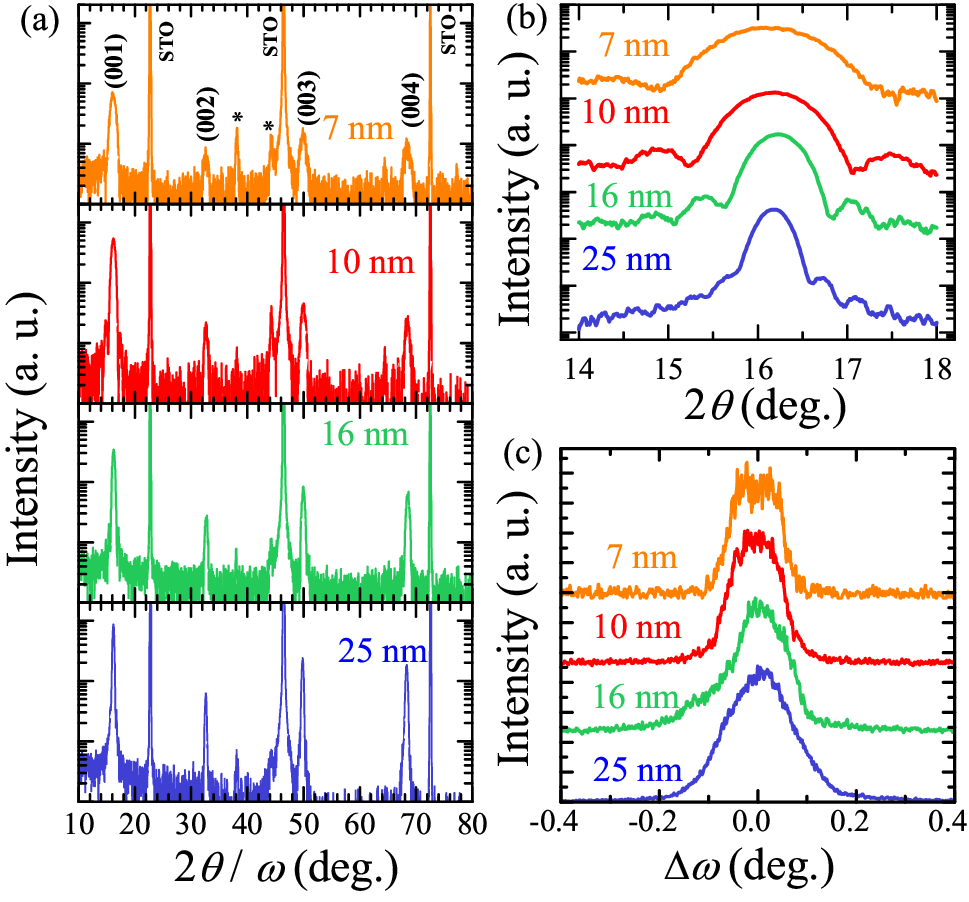}
\caption{(a) XRD patterns of the grown films. Asterisk represents reflections from Ag paste. (b) Enlarged plot around the (001) reflections. (c) Rocking curves of the (001) peaks of the grown films.}
\label{fig:XRD_result} 
\end{figure}

Figure \ref{fig:XRD_result}(a) illustrates the XRD patterns of four representative films with different thicknesses.
All the films indicated only the (00\textit{l}) reflection peaks of tetragonal FeSe, except for the peaks derived from the substrates and Ag paste attached to the films for the resistivity measurement.
These results indicate the single-phase nature of FeSe and  \textit{c}-axis orientation of the films.
In addition, we confirmed the in-plane alignment of the films as FeSe [100]/$\!$/STO [100].
Figure \ref{fig:XRD_result}(b) illustrates an enlarged plot of the (001) reflection peaks.
All the films showed clear Laue fringes, indicating a flat surface.
Figure \ref{fig:XRD_result}(c) illustrates the rocking curves of the (001) reflection peaks of the films.
All films showed a full width at half maximum (FWHM) smaller than 0.16\(^\circ\), indicating high crystalline quality.
All the films were under tensile strain with their \textit{a}-axis constant larger than \add{3.79} \AA \ (cf. bulk FeSe has $a$ = 3.77 \AA).
The structural properties of the films are summarized in table \ref{table:property}.

\begin{figure*}[hbt]
\includegraphics[width=\linewidth]{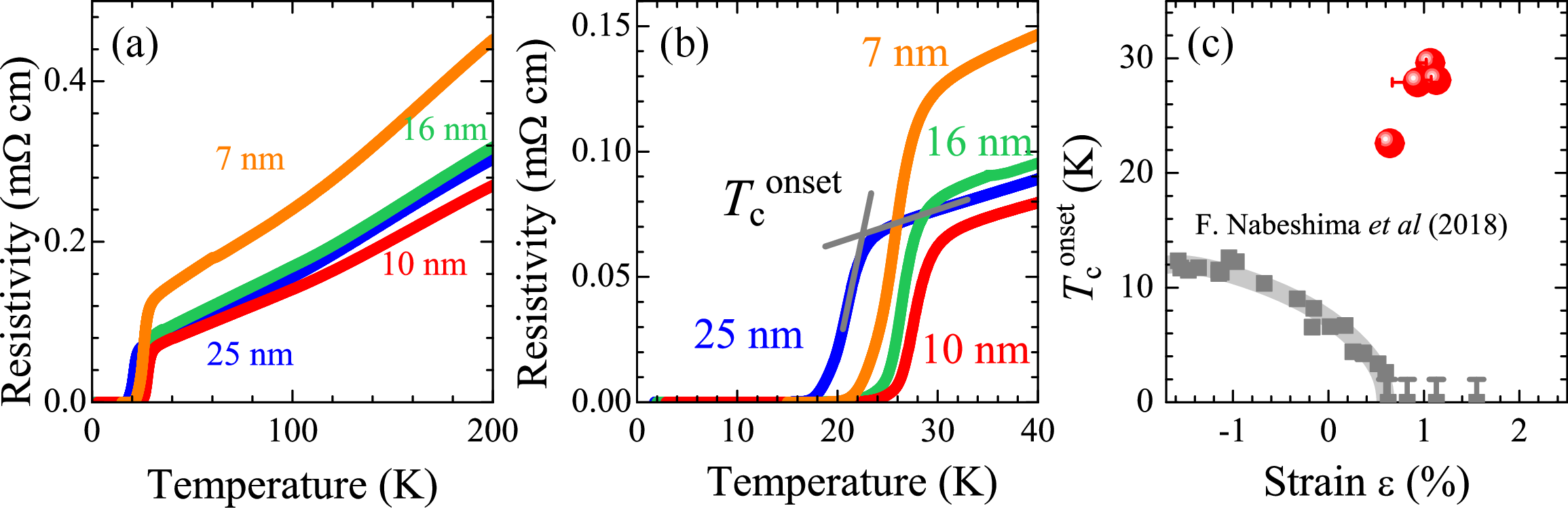}
\caption{(a), (b) Temperature dependence of the resistivity of the grown films.
(c) \Tconset as a function of strain of the films together with the data in \cite{Nabeshima_2018}.} 
\label{fig:result} 
\end{figure*}
\begin{table*}[htbp]
\caption{\label{table:property} Characteristics of the grown films. \textit{d} represents the film thickness.}
\begin{indented}
\item[]\begin{tabular}{cccccc}
\br
Sample &\textit{d} (nm)&  $a$ (\AA) &  $c$ (\AA) & \Tconset (K)& \Tczero (K)\\
\mr
\#1& 7 &\add{$3.808 \pm 0.009$}&$5.48\add{5 \pm 0.006}$&28 & 17\\
\#2& 10 &\add{$3.814 \pm 0.002$}&$5.47\add{2 \pm 0.002}$& 30 &19\\
\#3& 16 &\add{$3.816 \pm 0.002$}&$5.46\add{7 \pm 0.002}$& 28 & 17\\
\#4& 25 &{\add{$3.794 \pm 0.004$}}&$5.48\add{3 \pm  0.003}$& 23 & 13\\
\br
\end{tabular}
\end{indented}
\end{table*}

Figures \ref{fig:result}(a) and (b) illustrate the temperature dependence of resistivity $\rho(T)$ of the films.
All films exhibited a metallic behavior. 
Remarkably, all samples exhibited a superconducting transition with a \Tconset higher than 20 K.
These \Tconset values were much higher than those of the bulk at ambient pressure.

It is well known that in-plane strain strongly affects the superconducting properties of FeSe \cite{Feng_2018,Nabeshima_2018,Ghini_2021,Nakajima_2021}.
FeSe thin films under compressive strain exhibit enhanced \Tconset up to 12 K when $\epsilon \equiv(a_\mathrm{film}-a_\mathrm{bulk})/a_\mathrm{bulk} < -1.0\%$ \cite{Nabeshima_2018}, where $a_\mathrm{film}$ and $a_\mathrm{bulk}$ are the \textit{a}-axis constants of the films and bulk \cite{McQueen_2009}, respectively. 
However, tensile strain decreases \Tc and completely suppresses superconductivity when $\epsilon$ is greater than +0.6\% \cite{Nabeshima_2018}.
Similar results were obtained in subsequent studies on bulk crystals \cite{Nakajima_2021, Ghini_2021}.
We plot \Tconset of the films grown in this study in the strain-vs-\Tconset plot \cite{Nabeshima_2018} in figure \ref{fig:result}(c).
The present data largely deviate from the well-established strain-\Tc relation. 
The $\epsilon$ values in the grown films, which are higher than \add{0.6}$\%$, should lead to non-superconductivity, according to the strain-\Tc relation. 
Thus, these high \Tc values cannot be explained by the established strain effect.
This result suggests that interface effects are responsible for the high \Tc of the films.
\begin{figure}[hbt]
\includegraphics[width=\linewidth]{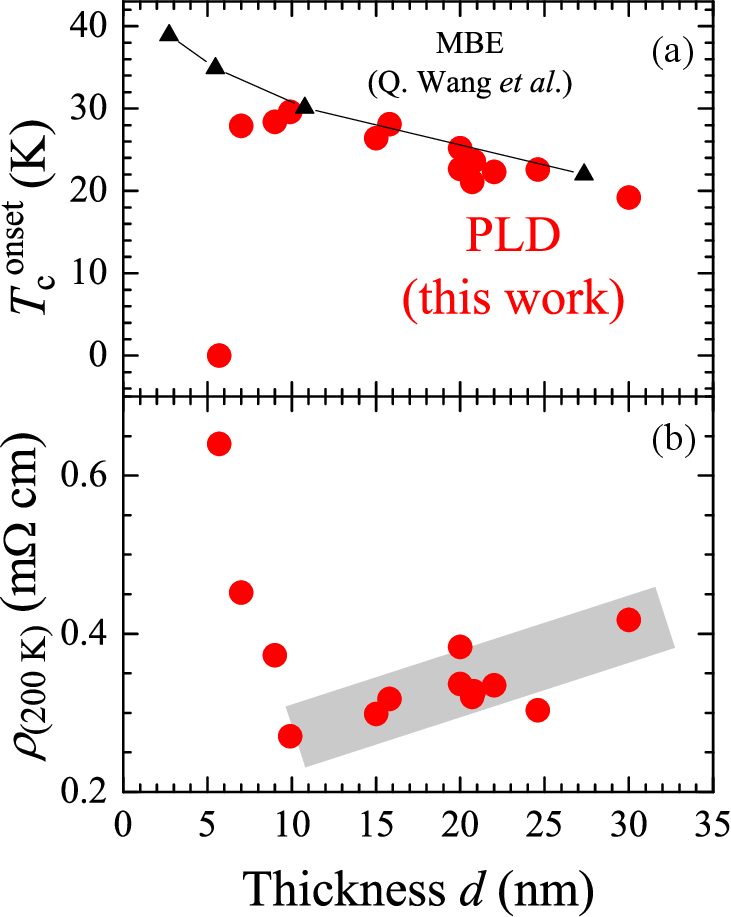}
\caption{(a) Thickness dependence of \Tconset of the films. 
Data for the films grown by MBE \cite{Wang_2015} are also plotted (black triangles). (b) Thickness dependence of resistivity at 200 K of the films.} 
\label{fig:figure4} 
\end{figure}

The interface superconductivity in the grown films is also suggested by the thickness dependence of \Tconset$\!$.
Figure \ref{fig:figure4}(a) shows \Tconset as a function of film thickness \textit{d} for all the films grown in this study. In the case of films with \textit{d} $\geq$ 10 nm, \Tconset increases as the thickness decreases.
This negative correlation between the film thickness and \Tc is in good agreement with the results obtained for MBE-grown FeSe/STO, which exhibited interface superconductivity \cite{Wang_2015,Zhao_2018}.
The negative correlation between film thickness and \Tc was interpreted as follows \cite{Wang_2015}: Because FeSe layers apart from the interface are considered not to be subject to interfacial effects \cite{Wang_2012}, it is likely that the very thin layer in the vicinity of the interface becomes superconducting.
When the thickness of the film is increased, the number of doped electrons at the superconductive interface of FeSe/STO is reduced, which results in a reduction of $T_{\mathrm{c}}$.
Although the validity of this interpretation is still open, the results of the thickness dependence, together with the extraordinarily high \Tc in terms of the strain-\Tc relation indicate the interface superconductivity in the films grown on STO. 
Note that the thickness dependence of \Tc is in sharp contrast to that of FeSe thin flakes \cite{Farrar_2020,Zhu_2021} and FeSe on bilayer graphens \cite{Song_2011} , where \Tc decreases as the thickness decreases.
This difference is consistent with the fact that the interface between FeSe and STO plays a crucial role in realizing enhanced superconductivity in the grown films.

Compared with MBE-grown films \cite{Wang_2015,Zhao_2018}, all films with \textit{d} $\geq$ 10 nm exhibited approximately the same value of \Tconset (figure \ref{fig:figure4}(a)).
Remarkably, our films exhibited a high \Tc without post-annealing.
MBE growth requires a post-annealing process to remove excess Se \cite{Zhang_2014} which suppresses the superconductivity.
The excess Se is due to the Se-rich growth conditions of the MBE techniques.
However, approximately the same composition is transferred from the Fe$_{1.1}$Se target to a film during PLD growth, which results in much less excess Se in the grown film.
This feature of the PLD technique enables us to obtain a superconducting FeSe/STO without post-annealing, which is different from MBE.

The PLD-grown films with \textit{d} $<$ 10 nm did not show a significant increase in \Tconset as the film thickness decreased, different from the MBE-grown films, and the 5-nm-thick film did not even exhibit superconductivity.

Figure \ref{fig:figure4}(b) illustrates resistivity at 200 K for all samples. 
For \textit{d} $\geq$ 10 nm, resistivity tends to decrease as the thickness decreases, possibly due to the existence of carrier-doped interfaces. However, for \textit{d} $<$ 10 nm, the resistivity rapidly increases as the film becomes thinner.
One possible explanation for the thickness dependence of the resistivity is degradation by air exposure (see supplementary materials). 
As film thickness decreases, FeSe layer near the interface may deteriorate more immediately, which may affect \Tc$\!$.
Thus, deposition of a protection layer on FeSe/STO is necessary for films thinner than 10 nm to obtain a higher \Tc$\!$, which is now under way.

\section{Conclusion}

In conclusion, we fabricated FeSe films on STO with a thickness of 5--30 nm which exhibited superconductivity with \Tconset much higher than that of bulk FeSe under ambient pressure.
The observed \Tc in the grown films was exceptionally high in terms of the strain-\Tc  relation of the same material established so far.
\Tc increased with decreasing film thickness, except for films thinner than 10 nm. This thickess dependence of \Tc is in good agreement with that of MBE-grown films that exhibited interface superconductivity \cite{Wang_2015, Zhao_2018}.
These results indicate that we successfully realized interface superconductivity in PLD-grown FeSe/STO. 
Furthermore, the PLD growth method has the advantage of realizing interface superconductivity without post-annealing, which is different from MBE techniques. 
PLD has the advantage of fabricating various heterostructures because different materials can be deposited by simply changing the target materials.
Thus, our results accelerate a study of interface superconductivity toward higher \Tc by systematically controlling the interface.

\begin{ack}
We would like to acknowledge K. Ueno and H. Okuma at the University of Tokyo for their technical support of the XRD and AFM measurements. \add{We also thank A. Ichinose at CRIEPI for the cross-sectional TEM experiment. }
\end{ack}

\section*{References}
\bibliography{FeSe_PLD_arxiv3}
\bibliographystyle{iopart-num}
\end{document}